# Low temperature reflectivity study of nonpolar ZnO/(Zn,Mg)O quantum wells grown on M-plane ZnO substrates

L. Béaur[1,2], T. Bretagnon[1,2], C. Brimont[1,2], T. Guillet[1,2], and B. Gil[1,2],

[1] Université Montpellier 2, Laboratoire Charles Coulomb, UMR5221, F-34095 Montpellier, France

[2] CNRS, Laboratoire Charles Coulomb, UMR5221, F-34095 Montpellier, France

D. Tainoff[3], M. Teisseire[3] and J.-M. Chauveau[3,4]

[3] CRHEA–CNRS, Rue Bernard Grégory, F-06560 Valbonne, France

[4] University of Nice Sophia Antipolis, Physics Dept., Parc Valrose, F-06102 Nice Cedex 2, France

PACS: 78.67.De ; 78.66.Hf ; 71.35.-y

We report growth of high quality ZnO/Zn$_{0.8}$Mg$_{0.2}$O quantum wells on M-plane oriented ZnO substrates. The optical properties are studied by reflectance spectroscopy. The optical spectra reveal strong in-plane optical anisotropies, as predicted by group theory, and clear reflectance structures, as an evidence of good interface morphologies. Signatures of confined excitons built from spin-orbit split-off valence band, analogous to C-exciton in bulk ZnO, are detected using a light polarized along the c-axis. Experiments performed in orthogonal polarization, show confined states analogous to A and B bulk excitons. Envelope function calculations including excitonic interaction nicely match the experimental results.

Wurtzite semiconductor heterostructures have been the subject of intense investigations during the last two decades, in connection with the potential application of these materials [1, 2] for optoelectronic applications from the visible to the ultraviolet region of the electromagnetic spectrum. However, such structures grown along the c-axis exhibit a Quantum Confined Stark Effect (QCSE) due to the large internal electric field along this direction. This is the case for polar ZnO heterostructures [3-6]. In order to eliminate the effect of the built-in electrical field on the optical properties, one can grow them along non-polar orientation [7-10]. In this letter we report reflectance investigations performed on a series of ZnO/(Zn,Mg)O single quantum wells grown on M-plane oriented ZnO substrates. The observation of both ground state excitons and spin-orbit split-off excited state ones demonstrates a reasonably small interface roughness, evidence of a highly controlled growth procedure.

The samples have identical designs and all consist of a bulk ZnO crystal carefully prepared so that its (10-10) plane can be used as a growth plane. A 200 nm thick (Zn,Mg)O barrier layer is first deposited, then a ZnO well layer, and finally a second (Zn,Mg)O barrier layer. The details of the growth procedure of the quantum wells can be found elsewhere [10]. Growth along the M-plane orientation leads to a strong anisotropy of the optical properties: in normal incidence reflectivity experiments, $\Gamma_1$ excitons can be probed using a light polarized along the [0001] direction and $\Gamma_5$ excitons are revealed for light polarized orthogonally to this direction. A detailed description of reflectivity properties of ZnO layers, including birefringence or polariton fine structure effects is out of the scope of this letter but can be found in Ref. 11. We wish to recall here that, thanks to the large value of the crystal field splitting in ZnO, the topmost valence band states are built from $p_x$ and $p_y$ type wave functions (in a spinless description) whilst the crystal field split one is mainly built from $p_z$ type states [12]. Thus, for a reflectivity experiment performed under normal incidence conditions, and for a electric field of the light parallel to the [0001] direction, the most intense optical transition will be the one associated with excitons built from states at the conduction band bottom and hole states built from this excited state hole. The corresponding excitonic transition (C line) is very strong in these experimental conditions. In contrast, in crossed polarization configuration, most of the oscillator strength is detected at the lower energies of the A and B excitons (excitons built from states at the conduction band bottom and hole states built from the topmost valence band states). The intensity of line A is smaller than the intensity of line B which dominates the spectrum. The transition energies of these lines are typically 3.376 eV and 3.383 eV for $\Gamma_{5T}$ A and B excitons, 3.425 eV for $\Gamma_1$ C exciton at liquid helium temperature. In ZnO quantum wells grown on a (10-10) oriented surface, electrons and holes are confined. Even in the case of a homoepitaxial quantum well the degeneracy of the $p_x$ and $p_y$ states in the bulk is lifted, due to the breaking of translational symmetry along the growth direction.. The optical response is anisotropic in this orthorhombic symmetry growth plane.

All of the following reflectance results were obtain at 10K using a Xenon lamp and a Jobin-Yvon iHR550 spectrometer. Figure 1 summarizes a series of reflectivity experiments carried out in normal incidence conditions. The light is polarized parallel to the [0001] direction, thus $\Gamma_1$ excitons are probed by this experiment. A strong signature of the 1s C exciton together with a weaker signature of the 2s state are detected in the bulk at low and higher energy respectively. Note the phase differences between the reflectance features of the samples due to slightly different thickness of the top barrier layer through the series of samples. The bulk exciton always occurs at the same energy and its full width at half maximum does not vary from sample to sample. Most interesting is the observation of a high energy feature, typical of these quantum wells, which blue shifts when the thickness of the ZnO quantum well layer decreases. This structure is marked with vertical arrows. The inhomogeneous broadening increases with decreasing the well width, due to the increase of the penetration of the wave functions into the barrier layers and the increased sensitivity of the transition energy to the roughness of the heterointerfaces. In case of the thinnest well this transition could no longer be detected. The energy of this transition decreases and reaches asymptotically the one of the bulk ZnO as the quantum well thickness increase. This is an evidence of the absence of QCSE in these samples, as expected from group theory [13].

Figure 2 summarizes a series of reflectivity experiments carried out in normal incidence conditions. The electric field of the photon is now orthogonal to the [0001] direction. The bulk $\Gamma_5$ excitons and their confined analogs are therefore probed by this experiment. Indeed, strong signatures of the $\Gamma_5$ A and $\Gamma_5$ B bulk excitons and their 2s excited states are detected. Here again, note the phase differences from sample to sample. The energies of bulk excitons and their full width at half maximum are not sample-dependent. The high energy feature blue shifts when the thickness of the ZnO quantum well layer decreases, yet its full width at half maximum prevents us from detecting A-B splitting. Again, the energies of these quantum confined excitons decrease with increasing well width but stay above that of the ZnO substrate, as an evidence of the absence of QCSE in these samples.

The evolution of the transition energies with well width is plotted in figure 3. The trend is similar to the one earlier reported in homoepitaxial ZnO/(Zn,Mg)O quantum wells grown on A-plane oriented substrates [10]. To avoid bringing unnecessary complexity in the figure, we omitted to plot the 2s state of the bulk A, B and C lines whose energies overlap the energy of the unresolved (A, B) doublet and C confined states. The data have been compared with a one band envelope function calculations. This approximation is reasonable because the inhomogeneous broadening of the excitonic lines prevents the observation of A-B splitting in our quantum wells. Moreover, the knowledge of the electronic structure of (Zn,Mg)O is still very limited. The valence band orderings were taken identical in ZnO and (Zn,Mg)O [14]. The fundamental band gap of (Zn,Mg)O is set to 3835 meV at 10 K. The conduction-band offset is taken to be 80% of the total band-gap difference between well and barrier materials. The effective masses are taken isotropic: 0.28 for the electron and 0.59 for the A, B and C holes in units of the free electron mass [15]. Both the calculation and the experiment are plotted in figure 3. The band to band envelope function calculation was corrected by the evolution of the long range electron-hole Coulomb interaction computed in a one-band variational calculation. The excitonic transitions are labeled here as $(e_1,X_1)$, where 1 indicates the confined quantum number and X refers to the symmetry of the valence band of the bulk (X = A, B and C). The evolution of the excitonic binding energy versus the quantum well width for both C-plane oriented and M-plane oriented quantum wells is shown in figure 4 in the specific case of A and B excitons. In nonpolar quantum wells, one may note that this long-range interaction is not reduced by the Quantum Confined Stark Effect as expected in GaN/(Ga,Al)N C-plane grown quantum wells [16]. The variational calculation indicates that the exciton binding energy varies from 70 to 90 meV in these samples. This 10-30 meV enhancement with respect to the bulk value (60 meV) is slightly larger than the inhomogeneous broadening making this correction to the band to band calculation necessary.

In conclusion, we have reported the observation of polarized excitonic structures in a series of homoepitaxial ZnO/(Zn,Mg)O quantum wells grown on M-plane bulk ZnO substrates. The evolution of the quantum confined exciton indicates the absence of Quantum Confined Stark Effect. In the limit of wide wells, the splitting between anisotropically confined excitons asymptotically reaches the value of this splitting in the bulk. The interface roughness deduced from the inhomogeneous broadening of the reflectivity features does not indicate noticeable differences compared with quantum wells grown along the polar orientation, a definite advantage if optoelectronic applications are targeted.

**Acknowledgement:**

The authors acknowledge financial support of ANR under "ZOOM" project (Grant No. ANR-06-BLAN-0135), "DefiZnO" (ANR-09-MAPR-009) project and the collaboration project CEA/CNRS number C12899/047588.

**Figure captions**

**Figure 1 (color online):** Normal incidence reflectivity spectra of a series of M-plane oriented ZnO/Zn$_{0.8}$Mg$_{0.2}$O single quantum wells. The light electric field is polarized parallel to the [0001] direction of the crystal. Note the observation of 1s and 2s states of the C exciton in the bulk and the observation of confined excitons (transitions labeled e$_1$-C$_1$) at the energies marked by arrows.

**Figure 2 (color online):** Normal incidence reflectivity spectra of a series of M- plane oriented ZnO/Zn$_{0.8}$Mg$_{0.2}$O single quantum wells. The light electric field is polarized orthogonal to the [0001] direction of the crystal. Note the observation of 1s of A exciton and 1s and 2s states of the B exciton in the bulk. The confined excitons (transitions labeled e$_1$-(A,B)$_1$) are also observed at the energies marked by arrows. In this case the A-B splitting is not resolved experimentally.

**Figure 3 (color online):** Transition energies plotted versus well width for a series of M-plane oriented ZnO/Zn$_{0.8}$Mg$_{0.2}$O single quantum wells. The energies of the excitons in the bulk ZnO substrate are indicated using open symbols. The full square (red) symbols represent energies of e$_1$(A,B)$_1$ lines whilst full circle (blue) ones correspond to e$_1$-C$_1$ confined excitons. The result of the effective mass envelope function calculation, including hole-electron interactions, is represented by black lines.

**Figure 4 (color online):** Result of the one parameter variational calculation of the exciton binding energies for M-plane (green) and C-plane (red) grown ZnO-

$Zn_{0.8}Mg_{0.2}O$ single quantum wells. QCSE is included in the case of the C-plane orientation; the internal electric field is taken as 0.82 MV/cm in the calculation.

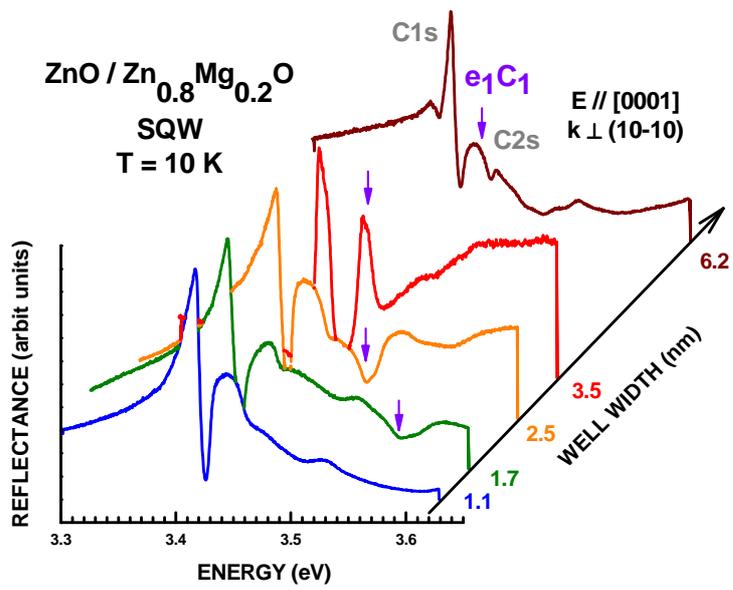

Figure 1

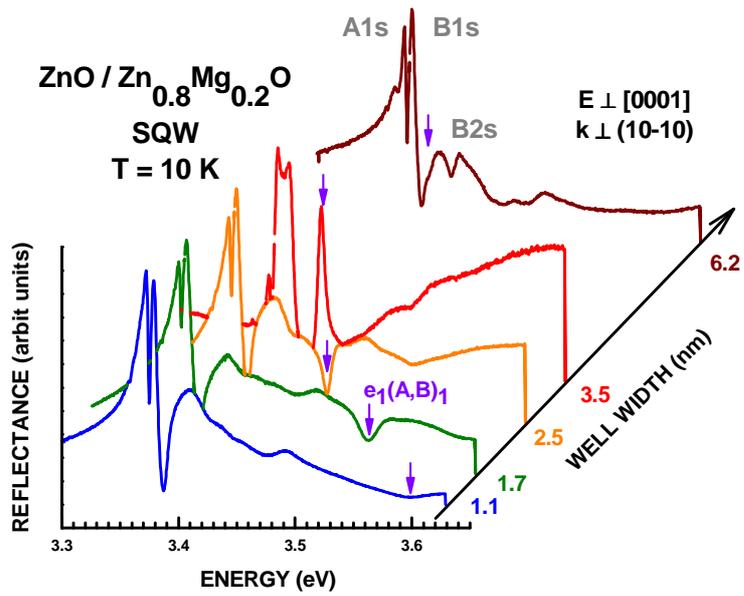

Figure 2
11

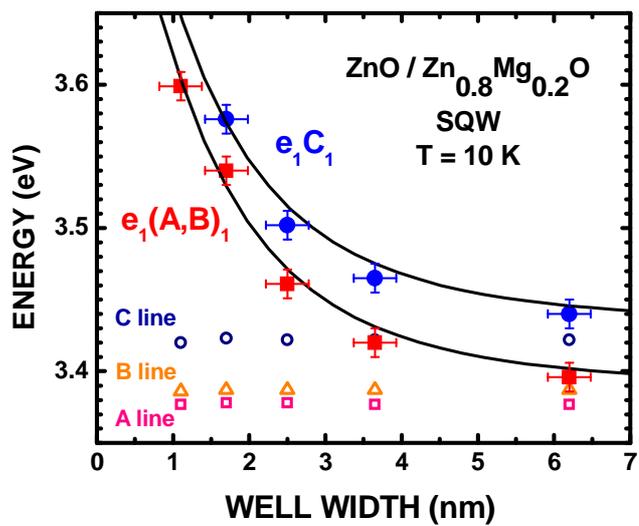

Figure 3

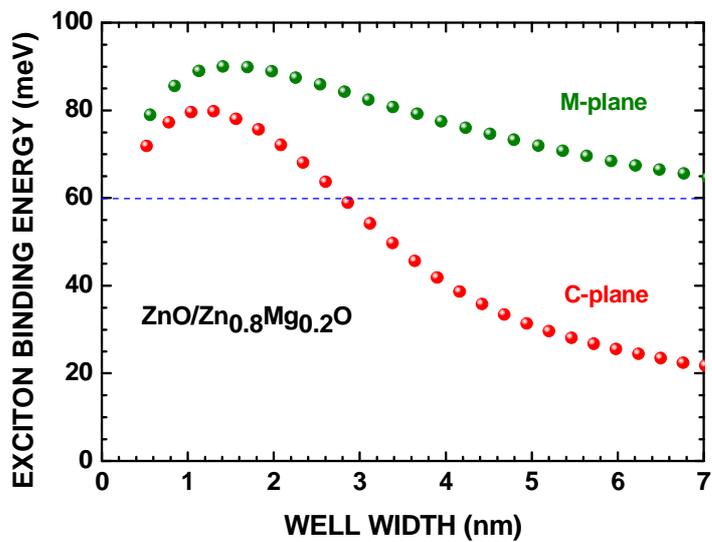

Figure 4